\documentclass[twocolumn,american,aps,prb,superscriptaddress]{revtex4-2}
\usepackage[T1]{fontenc}
\usepackage[utf8]{inputenc}
\usepackage{units}
\usepackage{amsmath}
\usepackage{amssymb}
\usepackage{graphicx}

\makeatletter
\usepackage{babel}

\makeatother

\usepackage{babel}
\begin{document}
\title{Controlling Ultrafast Excitations in Germanium:\\
The Role of Pump-Pulse Parameters and Multi-Photon Resonances}
\author{Amir Eskandari-asl}
\affiliation{Dipartimento di Fisica ``E.R. Caianiello'', Università degli Studi
di Salerno, I-84084 Fisciano (SA), Italy}
\author{Adolfo Avella}
\affiliation{Dipartimento di Fisica ``E.R. Caianiello'', Università degli Studi
di Salerno, I-84084 Fisciano (SA), Italy}
\affiliation{CNR-SPIN, Unità di Salerno, I-84084 Fisciano (SA), Italy}
\affiliation{CNISM, Unità di Salerno, Università degli Studi di Salerno, I-84084
Fisciano (SA), Italy}
\begin{abstract}
We employ the Dynamical Projective Operatorial Approach (DPOA) to
investigate the ultrafast optical excitations of germanium under intense,
ultrashort pump pulses. The method has very low resource demand relative
to many other available approaches and enables detailed calculation
of the residual electron and hole populations induced by the pump
pulse. It provides direct access to the energy distribution of excited
carriers and to the total energy transferred to the system. By decomposing
the response into contributions from different multi-photon resonant
processes, we systematically study the dependence of excited-carrier
density and absorbed energy on key pump-pulse parameters: duration,
amplitude, and photon energy. Our results reveal a complex interplay
between these parameters, governed by resonant Rabi-like dynamics
and competition between different multi-photon absorption channels.
For the studied germanium setup, we find that two-photon processes
are generally dominant, while one- and three-photon channels become
significant under specific conditions of pump-pulse frequency, duration,
and intensity. This comprehensive analysis offers practical insights
for optimizing ultrafast optical control in semiconductors by targeting
specific multi-photon pathways.
\end{abstract}
\maketitle

\section{Introduction}

In recent years, ultrafast pump--probe spectroscopy has emerged as
a pivotal experimental advancement, enabling unprecedented access
to the dynamics of condensed matter systems on femtosecond and even
attosecond time scales \cite{brabec2000intense,krausz2009attosecond,krausz2014attosecond,calegari2016advances,gandolfi2017emergent,borrego2022attosecond,inzani2023field,inzani2023photoinduced}.
By tracking the dynamics of photo-excited carriers, these experiments
provide direct insights into the microscopic mechanisms governing
electronic, spin, and lattice degrees of freedom far from equilibrium.
From a technological perspective, such understanding is essential
for the development of next-generation ultrafast optoelectronic and
spintronic devices. Fundamentally, these methods allow for the real-time
observation of symmetry breaking, coherence, and relaxation processes
\cite{Zurch_17,PhysRevB.97.205202,perfetti2008femtosecond}.

To theoretically address the complexity of ultrafast phenomena in
realistic materials, we have recently developed the Dynamical Projective
Operatorial Approach (DPOA) \cite{inzani2023field,eskandari2024time,eskandari2024generalized,eskandari2024out,eskandari2025dynamical}.
DPOA is an efficient operator-based formalism that enables the simulation
of real-time evolution in multi-band systems under the application
of pump fields. It provides direct access to key microscopic observables,
such as the single-particle density matrix (SPDM) and band populations
\cite{eskandari2025controlling}, inter-band coherences, and ---
in principle --- every multi-time multi-particle response function,
including time-resolved angle-resolved photoemission spectroscopy
\cite{eskandari2024time} and transient optical properties \cite{eskandari2024generalized}.
The formulation of DPOA is fully general, applicable to systems with
arbitrary lattice structures, band numbers, and other complexities.

A central question in ultrafast optical excitations concerns the distribution
of excited electrons and holes in energy and momentum space. Specifically:
At which energies do the excited carriers reside? How many carriers
are excited under a given pump pulse? And how much energy is transferred
to the electronic system? Addressing these questions is crucial for
advancing toward practical applications of ultrafast physics, such
as optimizing pulse parameters for specific device functionalities.
In this work, we apply DPOA to study the excitations in germanium
driven by an intense ultrashort optical pulse, systematically analyzing
the dependence of residual excited-carrier density and absorbed energy
on pump-pulse parameters.

The manuscript is organized as follows. In Sec.~\ref{sec:Theory},
we present the theoretical framework, including the Hamiltonian in
the dipole gauge, the construction of time-dependent hopping and dipole
matrices via the Wannier basis, and the DPOA equations of motion.
We also define the key quantities of interest: residual carrier populations,
their photon-order decomposition, and the associated energy distributions.
In Sec.~\ref{sec:Numerical-studies}, we report numerical results
for germanium, examining the roles of pump-pulse duration, amplitude,
and frequency in determining the excitation pathways and efficiency.
Finally, in Sec.~\ref{sec:Conclusions}, we summarize our findings
and discuss their implications for future studies and applications.

\section{Theory}\label{sec:Theory}

In this section, we present the relevant aspects of the DPOA framework
for the analysis of residual excited-carrier populations and absorbed
energy. Moreover, we define the fundamental quantities that enable
us to conduct a comprehensive study of such excitations and achieve
a deeper understanding of their underlying processes. It is worth
noting that, in principle, this study is highly resource-intensive
and could render many other approaches infeasible. In contrast, the
extreme efficiency of the DPOA framework allows the required resources
to remain well contained, thereby enabling analysis of a large portion
of the relevant parameter space.

Working in the dipole gauge, the time-dependent Hamiltonian governing
a pumped system is given by \cite{schuler2021gauge,eskandari2024time,eskandari2024generalized}
\begin{equation}
\mathcal{H}(t)=\sum_{\mathbf{k},n,n'}c_{\mathbf{k},n}^{\dagger}(t)\Xi_{\mathbf{k},n,n'}(t)c_{\mathbf{k},n'}(t),
\end{equation}
where $c_{\mathbf{k},n}(t)$ annihilates an electron with crystal
momentum $\mathbf{k}$ and band index $n$ (including spin). The matrix
$\Xi_{\mathbf{k}}(t)$ originates from the first-quantized single-particle
Hamiltonian and takes the form 
\begin{equation}
\Xi_{\mathbf{k}}(t)=T_{\boldsymbol{k}}(t)+eE_{\text{pu}}(t)D_{\boldsymbol{k},\text{pu}}(t),\label{eq:ham_dg-b}
\end{equation}
with $D_{\boldsymbol{k},\text{pu}}(t)=\boldsymbol{D}_{\boldsymbol{k}}(t)\cdot\hat{\boldsymbol{u}}_{\text{pu}}$.
Here, $T_{\boldsymbol{k}}(t)$ and $\boldsymbol{D}_{\boldsymbol{k}}(t)$
represent the time-dependent hopping and dipole matrices, respectively.
The applied linearly polarized pump pulse is characterized by the
vector potential 
\begin{equation}
\boldsymbol{A}_{\text{pu}}(t)=A_{\text{pu}}(t)\hat{\boldsymbol{u}}_{\text{pu}},
\end{equation}
and the associated electric field 
\begin{equation}
\boldsymbol{E}_{\text{pu}}(t)=E_{\text{pu}}(t)\hat{\boldsymbol{u}}_{\text{pu}}=-\partial_{t}A_{\text{pu}}(t)\hat{\boldsymbol{u}}_{\text{pu}}.
\end{equation}
The unit vector $\hat{\boldsymbol{u}}_{\text{pu}}$ specifies the
pump-pulse polarization axis. The pump pulse is switched on from an
initial time $t_{\mathrm{ini}}\rightarrow-\infty$, when the system
resides in a thermal equilibrium state.

$T_{\boldsymbol{k}}(t)$ and $\boldsymbol{D}_{\boldsymbol{k}}(t)$
are built by implementing the Peierls substitution in a localized
Wannier basis and then rotated to the band basis that diagonalizes
the equilibrium Hamiltonian \cite{eskandari2024time,eskandari2024generalized}.
Operators expressed in the localized basis are hereafter marked with
a tilde ($\tilde{\;}$). The rotation from the localized to the band
basis is carried out by the unitary matrix $\Omega_{\mathbf{k}}$.
In compact notation, $\Omega_{\mathbf{k}}^{\dagger}\cdot\tilde{T}_{\mathbf{k}}\cdot\Omega_{\mathbf{k}}$
is diagonal, its entries being the band energies $\varepsilon_{\mathbf{k},n}$.
For any matrix $M$ that stands for either $T$ or $\boldsymbol{D}$,
we have 
\begin{equation}
M_{\mathbf{k}}(t)=\Omega_{\mathbf{k}}^{\dagger}\cdot\tilde{M}_{\mathbf{k}+\frac{e}{\hbar}\boldsymbol{A}_{\text{pu}}(t)}\cdot\Omega_{\mathbf{k}}.
\end{equation}
The dot symbol $\cdot$ is used for both vector products in real space
and matrix products in the electronic Hilbert space. In practical
calculations on multi-band systems with dense $\mathbf{k}$-meshes,
it is numerically efficient to expand $\tilde{M}_{\mathbf{k}+\frac{e}{\hbar}\boldsymbol{A}_{\text{pu}}(t)}$
as a power series in the vector potential through the Peierls expansion
\cite{eskandari2024time}.

Within the DPOA formalism, the Heisenberg-picture annihilation operators
$c_{\mathbf{k}}(t)$ are related to their equilibrium values $c_{\mathbf{k}}(t_{\mathrm{ini}})$
via the projection matrices $P_{\mathbf{k}}(t)$: 
\begin{equation}
c_{\mathbf{k}}(t)=P_{\mathbf{k}}(t)\cdot c_{\mathbf{k}}(t_{\mathrm{ini}}).
\end{equation}
Here, $c_{\mathbf{k}}(t)$ is a column vector whose entries are the
operators $c_{\mathbf{k},n}(t)$. The projection matrices satisfy
the equation of motion \cite{eskandari2024time}
\begin{equation}
i\hbar\partial_{t}P_{\mathbf{k}}(t)=\Xi_{\mathbf{k}}(t)\cdot P_{\mathbf{k}}(t),\label{eq:EOM_P}
\end{equation}
which is solved numerically starting from the initial condition $P_{\mathbf{k}}(t_{\mathrm{ini}})=\boldsymbol{1}$.

The electronic population in band $n$ at momentum $\mathbf{k}$ and
time $t$ is given by 
\begin{equation}
N_{\mathbf{k},n}(t)=\langle c_{\mathbf{k},n}^{\dagger}(t)c_{\mathbf{k},n}(t)\rangle.
\end{equation}
Using the projection matrices, this can be expressed as 
\begin{equation}
N_{\mathbf{k},n}(t)=\sum_{n'}P_{\mathbf{k},n,n'}(t)f(\varepsilon_{\mathbf{k},n'})P_{\mathbf{k},n',n}^{\dagger}(t),\label{eq:N_P}
\end{equation}
where $f(\varepsilon)=[e^{\beta\left(\varepsilon-\mu\right)}+1]^{-1}$
is the Fermi distribution function, in which $\beta$ is the inverse
temperature and $\mu$ is the chemical potential.

We consider a Gaussian-enveloped pump pulse: 
\begin{equation}
A_{\mathrm{pu}}(t)=A_{0}e^{-(4\ln2)t^{2}/\tau_{\mathrm{pu}}^{2}}\sin(\omega_{\mathrm{pu}}t),\label{eq:Apu}
\end{equation}
where $\tau_{\mathrm{pu}}$ is the full width at half maximum (FWHM)
of the pulse (also referred to as the pump-pulse duration), and $\omega_{\mathrm{pu}}$
is the pump-pulse frequency. The quantity $\hbar\omega_{\mathrm{pu}}$
is dubbed the pump-pulse photon energy.

The residual electronic population after the pulse, $N_{\mathbf{k},n}^{\mathrm{res}}=N_{\mathbf{k},n}(t\rightarrow\infty)$,
differs from the equilibrium value, $f(\varepsilon_{\mathbf{k},n})$,
only for bands that are in an $l$-photon resonance ($l=1,2,3,\dots$)
with the pump pulse. Here, $t\rightarrow\infty$ denotes times after
the application of the pump pulse but before relaxation processes
set in; such processes occur on time scales of hundreds of femtoseconds
and are neglected in this work \cite{inzani2022field}. The residual
excited-carrier population is 
\begin{equation}
\Delta N_{\mathbf{k},n}^{\mathrm{res}}=N_{\mathbf{k},n}^{\mathrm{res}}-f(\varepsilon_{\mathbf{k},n}).
\end{equation}
Here, $\Delta N_{\mathbf{k},n}^{\mathrm{res}}>0$ corresponds to an
electron in a conduction band (CB), while $\Delta N_{\mathbf{k},n}^{\mathrm{res}}<0$
corresponds to a hole in a valence band (VB). We consider a gapped
semiconductor at low temperature, so that the VBs and CBs are well
separated.

The residual excited-carrier population per unit cell, which can be
referred to as the residual excited-carrier density, is 
\begin{equation}
\Delta N^{\mathrm{res}}=\frac{1}{M_{\mathrm{grid}}}\sum_{\mathbf{k},n_{C}}\Delta N_{\mathbf{k},n_{C}}^{\mathrm{res}},\label{eq:N_}
\end{equation}
where $n_{C}$ runs over all conduction bands and $M_{\mathrm{grid}}$
is the total number of $\mathbf{k}$-points in the numerical grid.

The strength of an $l$-photon resonance for a given gap energy $\varepsilon_{\mathrm{gap}}$
is given by \cite{eskandari2024time} 
\begin{align}
w_{l}(\varepsilon_{\mathrm{gap}})=\exp\left[-\frac{\tau_{\mathrm{pu}}^{2}}{8\ln2\hbar^{2}l}\left(\varepsilon_{\mathrm{gap}}-l\hbar\omega_{\mathrm{pu}}\right)^{2}\right].\label{eq:strenght_res_l}
\end{align}
This expression resembles the squared amplitude of the $\varepsilon_{\mathrm{gap}}/\hbar$
component in the spectrum of the $l$-th power of the pump pulse,
which is centered at $l\omega_{\mathrm{pu}}$.

The contribution of a specific VB, $n_{V}$, to the residual excited
electron population in a given CB, $n_{C}$, via an $l$-photon process
is estimated as \cite{eskandari2024time}
\begin{equation}
\Delta N_{\mathbf{k},n_{C},n_{V}}^{\mathrm{res}(l)}=\frac{\Delta N_{\mathbf{k},n_{V}}^{\mathrm{res}}w_{l}(\varepsilon_{\mathbf{k},n_{C}}-\varepsilon_{\mathbf{k},n_{V}})}{\sum_{n_{V}'}\Delta N_{\mathbf{k},n_{V}'}^{\mathrm{res}}\sum_{l'}w_{l'}(\varepsilon_{\mathbf{k},n_{C}}-\varepsilon_{\mathbf{k},n_{V}'})}\Delta N_{\mathbf{k},n_{C}}^{\mathrm{res}}.\label{eq:N_k_nc_nv_l}
\end{equation}
Summing over all VBs gives the total $l$-photon contribution to the
residual excited-carrier population residing in $n_{C}$: 
\begin{equation}
\Delta N_{\mathbf{k},n_{C}}^{\mathrm{res}(l)}=\sum_{n_{V}}\Delta N_{\mathbf{k},n_{C},n_{V}}^{\mathrm{res}(l)}.\label{eq:N_k_nc_l}
\end{equation}
The residual excited-carrier density via $l$-photon resonant processes
is 
\begin{equation}
\Delta N^{\mathrm{res}(l)}=\frac{1}{M_{\mathrm{grid}}}\sum_{\mathbf{k},n_{C}}\Delta N_{\mathbf{k},n_{C}}^{\mathrm{res}(l)}.\label{eq:N_l}
\end{equation}
Analogous expressions hold for hole residual excited-carrier populations
and densities, with the roles of CB and VB interchanged.

To resolve the energy distribution of the residual excited-carrier
density, we define 
\begin{equation}
\rho^{\mathrm{res}}(\varepsilon)=\frac{1}{M_{\mathrm{grid}}}\sum_{\mathbf{k},n}\Delta N_{\mathbf{k},n}^{\mathrm{res}}L(\varepsilon-\varepsilon_{\mathbf{k},n}),\label{eq:N_en}
\end{equation}
where $L(x)=\frac{1}{\pi}\frac{\lambda}{x^{2}+\lambda^{2}}$ is a
Lorentzian with damping factor $\lambda$, chosen according to the
$\mathbf{k}$-grid density \cite{eskandari2024generalized}. For an
infinitely fine grid, $\lambda\rightarrow0^{+}$ and hence, $L(x)\rightarrow\delta(x)$.
The $l$-photon resolved distribution is obtained by replacing $\Delta N_{\mathbf{k},n}^{\mathrm{res}}$
with $\Delta N_{\mathbf{k},n}^{\mathrm{res}(l)}$: 
\begin{equation}
\rho^{\mathrm{res}(l)}(\varepsilon)=\frac{1}{M_{\mathrm{grid}}}\sum_{\mathbf{k},n}\Delta N_{\mathbf{k},n}^{\mathrm{res}(l)}L(\varepsilon-\varepsilon_{\mathbf{k},n}).\label{eq:N_en-l}
\end{equation}

Finally, the residual energy absorbed by the system per unit cell
is 
\begin{equation}
\mathcal{E}^{\text{res}}=\frac{1}{M_{\mathrm{grid}}}\sum_{\mathbf{k},n}\Delta N_{\mathbf{k},n}^{\mathrm{res}}\varepsilon_{\mathbf{k},n},\label{eq:EnCell}
\end{equation}
which is independent of the energy origin due to particle conservation.
The $l$-photon contribution is 
\begin{equation}
\mathcal{E}^{\text{res}(l)}=\frac{1}{M_{\mathrm{grid}}}\sum_{\mathbf{k},n}\Delta N_{\mathbf{k},n}^{\mathrm{res}(l)}\varepsilon_{\mathbf{k},n}.\label{eq:EnCell-l}
\end{equation}

\section{Excitations in Germanium}\label{sec:Numerical-studies}

\begin{figure}
\centering{}\includegraphics[width=8cm]{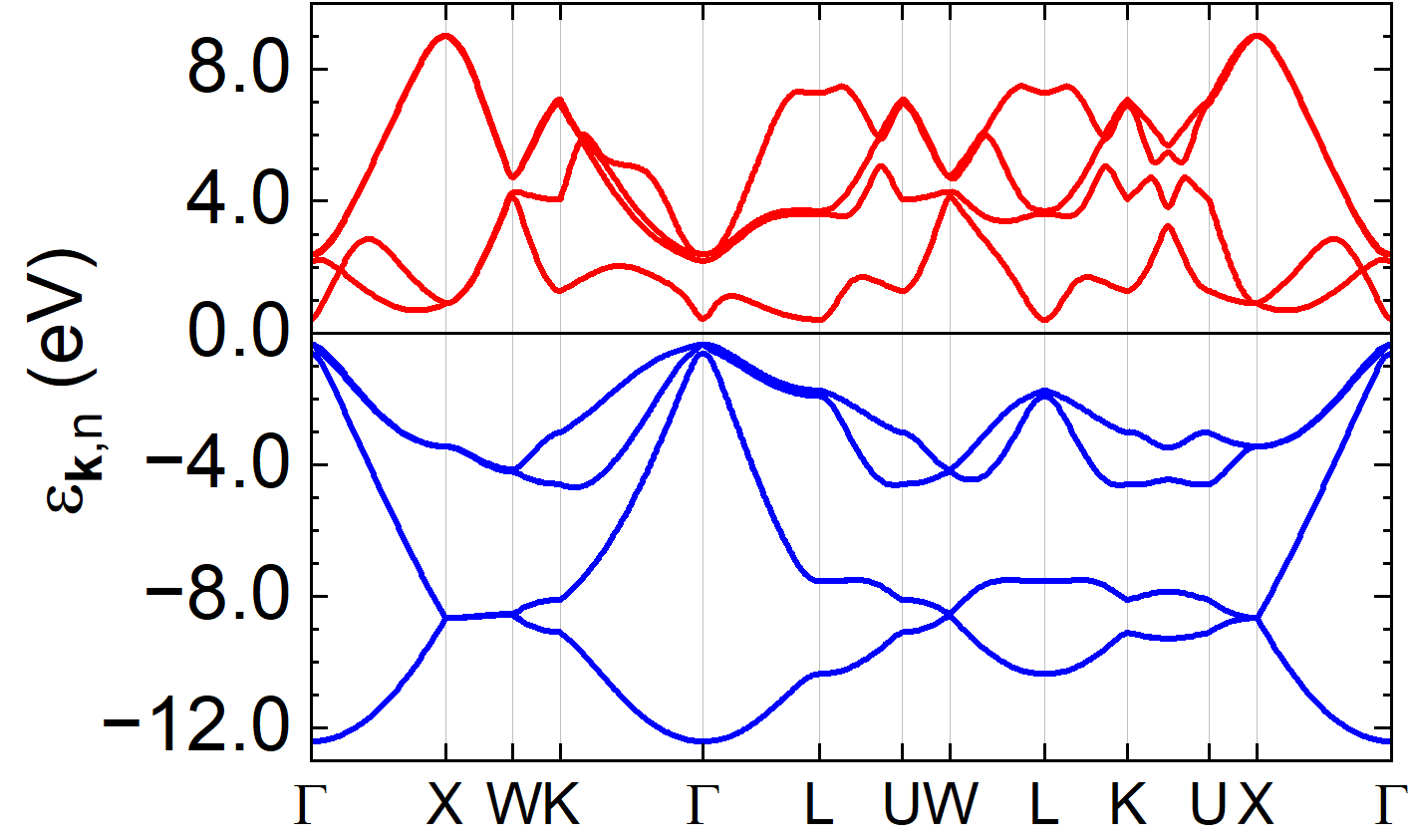}\caption{Equilibrium band structure of germanium along the main high-symmetry
lines.}\label{fig:bands}
\end{figure}

\begin{figure*}
\centering{}\includegraphics[width=18cm]{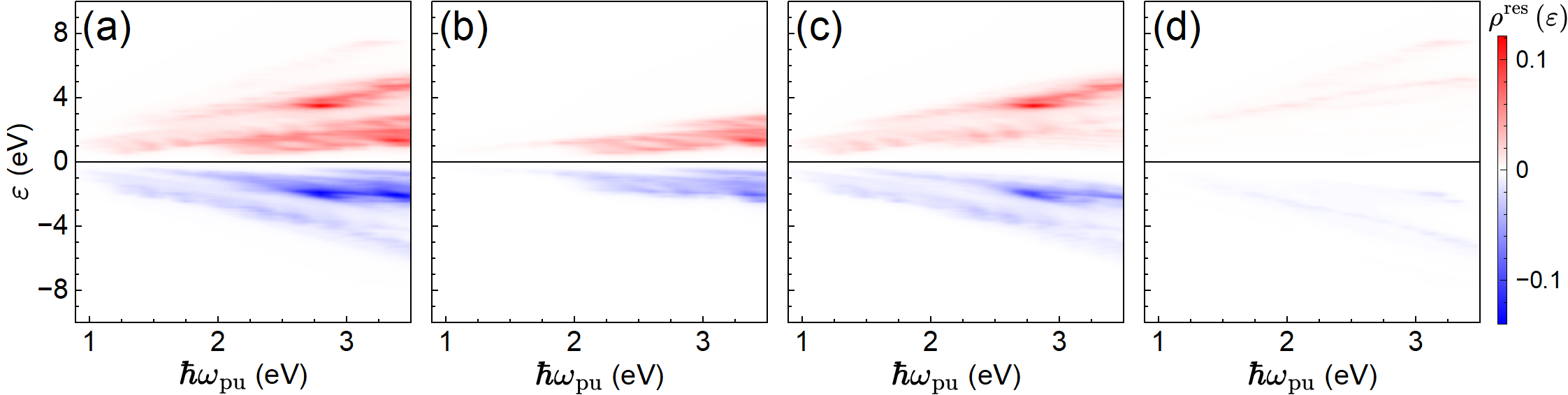}
\caption{Energy distribution of residual excited-carrier density, $\rho^{\mathrm{res}}(\varepsilon)$,
for different pump-pulse photon energies $\hbar\omega_{\mathrm{pu}}$.
(a) Total distribution. (b), (c), (d) Contributions from one-, two-,
and three-photon processes, respectively. Red (blue) color shows positive
(negative) excess populations and corresponds to electron (hole) excitations.
The vertical spacing between red and blue branches reflects the pump-pulse
photon energy multiples, indicating resonant origins.}\label{fig:rho}
\end{figure*}

In this section, we present numerical results for optically pumped
germanium, using the theoretical framework described in Sec.~\ref{sec:Theory}.
The electronic structure is obtained from first-principles calculations
performed with the Elk code \cite{elk_code}. The hopping parameters,
$\tilde{T}_{\mathbf{k}}$, and dipole matrix elements, $\tilde{\mathbf{D}}_{\mathbf{k}}$,
are computed using Wannier90 \cite{pizzi2020wannier90}, following
the procedure detailed in Ref.~\cite{inzani2023field}. We focus
on a closed manifold of 16 $sp^{3}$ bands around the chemical potential,
which captures the main band gap of approximately $\unit[800]{meV}$
at the $\Gamma$ point. The band structure, including spin-orbit coupling,
is shown in Fig.~\ref{fig:bands}. The band splitting due to spin-orbit
coupling is very small for the displayed bands. The chemical potential
is set to zero energy.

\begin{figure*}
\centering{}\includegraphics[width=18cm]{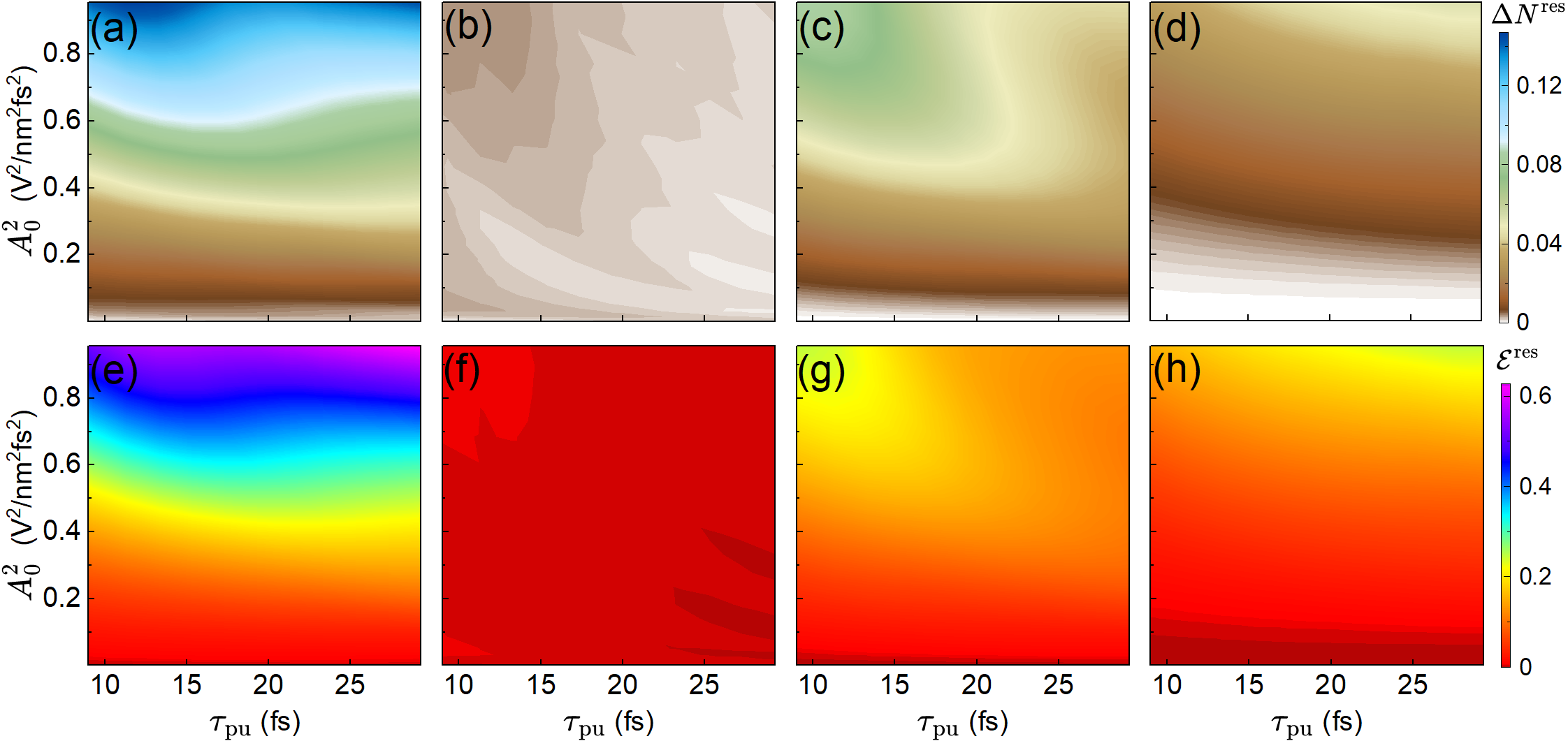}\caption{Residual excited-carrier density, $\Delta N^{\mathrm{res}}$, (upper
row) and residual absorbed energy per unit cell, $\mathcal{E}^{\mathrm{res}}$,
(lower row) as functions of pump-pulse duration, $\tau_{\mathrm{pu}}$,
and squared amplitude, $A_{0}^{2}$. (a, e) Total quantities. (b,
f), (c, g), (d, h) Contributions from one-, two-, and three-photon
processes, respectively.}\label{fig:A2-tau}
\end{figure*}

The crystal orientation and pump-pulse polarization are chosen as
in Ref.~\cite{inzani2023field}, with the field polarized along the
$[100]$ direction. The Brillouin zone is sampled on a $32\times32\times32$
$\mathbf{k}$-grid centered at $\Gamma$, and the Lorentzian broadening
in Eq.~\eqref{eq:N_en} is set to $\lambda=\unit[0.079]{eV}$ (equivalent
to $\unit[0.12]{PHz}$). Unless otherwise stated, the default values
of the pulse parameters are: $\tau_{\mathrm{pu}}=\unit[13.3]{fs}$,
$A_{0}=\unitfrac[0.528]{V}{nm\,fs}$, and $\hbar\omega_{\mathrm{pu}}=\unit[1.55]{eV}$.
In all subsequent parameter scans, any parameter not being varied
is held constant at its default value. Finally, the temperature is
set to zero in our calculations.

\begin{figure*}
\centering{}\includegraphics[width=18cm]{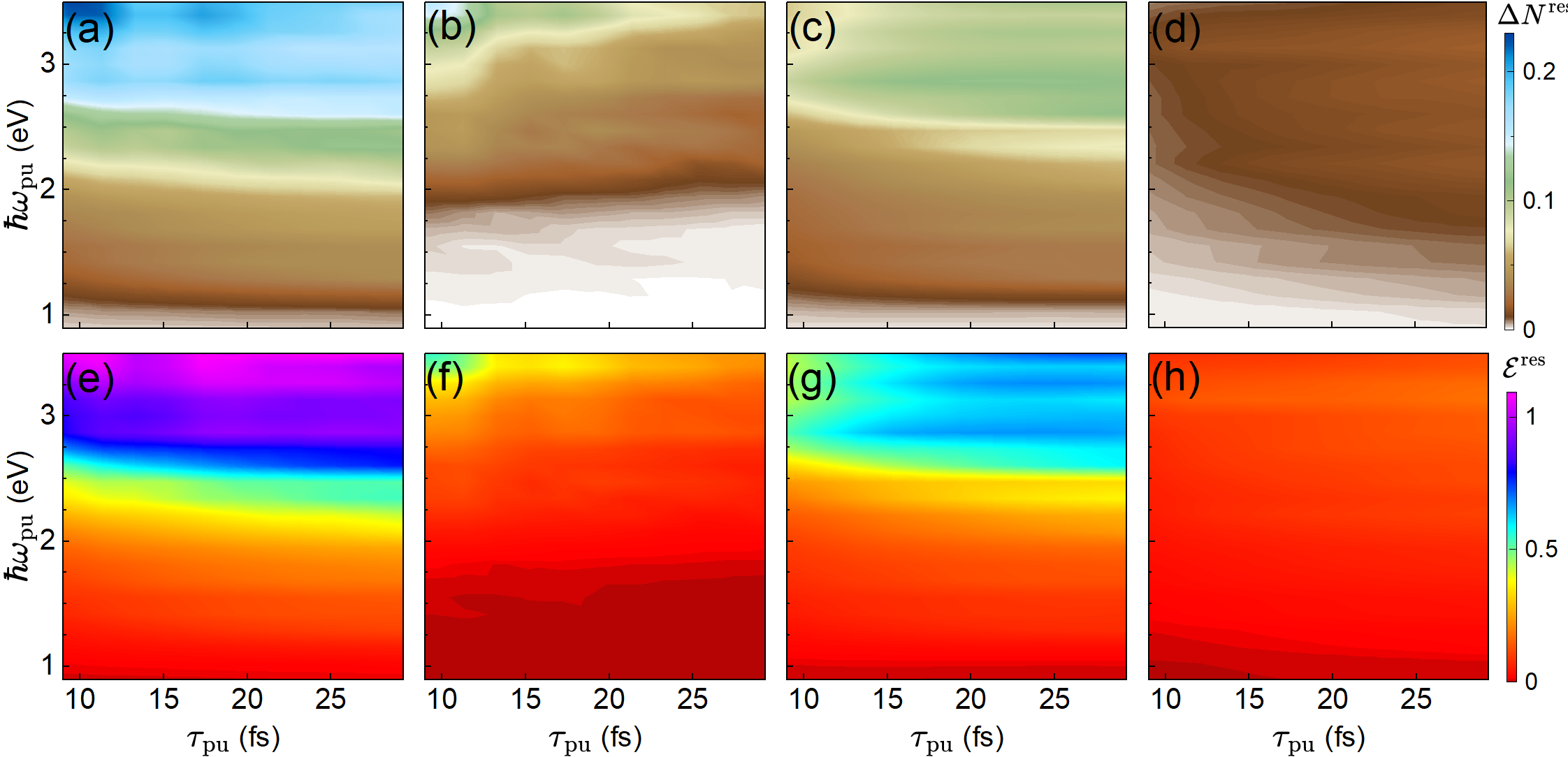}\caption{Residual excited-carrier density, $\Delta N^{\mathrm{res}}$, (upper
row) and residual absorbed energy, $\mathcal{E}^{\mathrm{res}}$,
(lower row) as functions of pump-pulse duration, $\tau_{\mathrm{pu}}$,
and pump-pulse photon energy, $\hbar\omega_{\mathrm{pu}}$. Columns
as in Fig.~\ref{fig:A2-tau}.}\label{fig:Enph-tau}
\end{figure*}

Figure~\ref{fig:rho}(a) displays the total energy distribution of
residual excited-carrier density, computed via Eq.~\eqref{eq:N_en}.
The positive (red) and negative (blue) branches correspond to electron
and hole excitations, respectively. Their separation in energy by
multiples of $\hbar\omega_{\mathrm{pu}}$ clearly indicates that the
excitations are due to $l$-photon resonant processes, where the resonance
strength is governed by Eq.~\eqref{eq:strenght_res_l}. The distributions
in panels (b)--(d) decompose the signal into one-, two-, and three-photon
contributions. Notably, for lower pump-pulse frequencies, two-photon
processes dominate, consistent with earlier findings \cite{inzani2023field}.
As the pump-pulse frequency increases, the role of one-photon excitations
grows. Three-photon contributions remain relatively small; higher-order
processes ($l>3$) are negligible and not shown here.

\begin{figure*}
\centering{}\includegraphics[width=18cm]{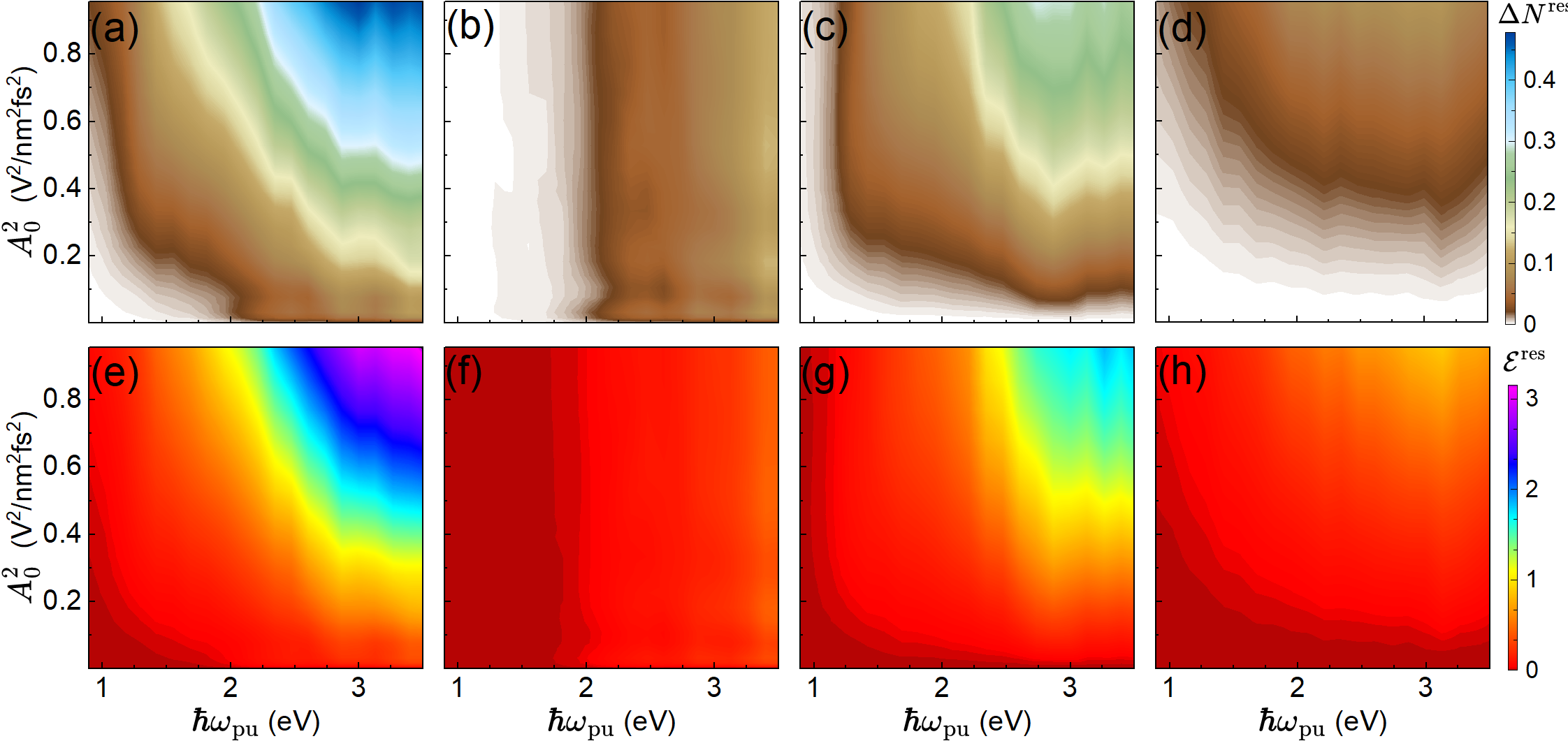}\caption{Residual excited-carrier density, $\Delta N^{\mathrm{res}}$, (upper
row) and residual absorbed energy per unit cell, $\mathcal{E}^{\mathrm{res}}$,
(lower row) as functions of pump-pulse photon energy $\hbar\omega_{\mathrm{pu}}$
and squared amplitude $A_{0}^{2}$. Columns as in Fig.~\ref{fig:A2-tau}.}\label{fig:A2-Enph}
\end{figure*}

Figure~\ref{fig:A2-tau} explores the dependence of the residual
excited-carrier density and absorbed energy per unit cell on pump-pulse
duration and amplitude. Panels (a) and (e) show that, for fixed $\tau_{\mathrm{pu}}$,
increasing $A_{0}$ generally enhances both $\Delta N^{\mathrm{res}}$
and $\mathcal{E}^{\mathrm{res}}$, as expected for higher pump-pulse
intensity. For small $A_{0}$, longer pulses monotonically increase
the excited population and absorbed energy, whereas for larger amplitudes
a non-monotonic Rabi-like behavior emerges: after an initial rise,
the residual excited-carrier density decreases as $\tau_{\mathrm{pu}}$
increases. This reflects the increased Rabi frequency at higher amplitudes,
which can drive populations back toward the ground state for sufficiently
long pulses.

The photon-resolved panels reveal distinct behaviors. One-photon processes
{[}panels (b, f){]} are generally weak. Moreover, they are suppressed
at longer $\tau_{\mathrm{pu}}$, especially at high amplitudes, where
Rabi oscillations are strong. Two-photon processes {[}panels (c, g){]}
dominate across most parameter ranges and persist for longer pulse
durations due to their smaller effective Rabi frequencies. Three-photon
contributions {[}panels (d, h){]} become significant only for high
amplitudes and long pulses, where they can even exceed the two-photon
signal. This sequential activation of higher-$l$ processes with increasing
$\tau_{\mathrm{pu}}$ stems from both differing Rabi frequencies and
inter-channel competition: once higher-order resonances are activated,
they reduce the available population for lower-order excitations.
This is possible because at a single $\mathbf{k}$-point we may have
several different resonances simultaneously.

A noteworthy observation is that in the regime where three-photon
processes dominate, their contribution in the absorbed energy per
unit cell is substantially higher than the corresponding contribution
of the two-photon resonances in their own regime, reflecting the larger
number of energy quanta involved in multi-photon absorption.

Figure~\ref{fig:Enph-tau} examines the joint dependence of the residual
excited-carrier density and absorbed energy per unit cell on $\tau_{\mathrm{pu}}$
and $\hbar\omega_{\mathrm{pu}}$. The total excitations {[}panels
(a, e){]} generally increase with $\hbar\omega_{\mathrm{pu}}$, but
exhibits pronounced dips and peaks that reflect the resonant matching
between the pump-pulse frequency and specific band gaps across the
Brillouin zone, as well as the variation of the strength of the coupling
to the pump pulse. The detailed $l$-photon decompositions {[}panels
(b--d, f--h){]} show that each photon order has a distinct frequency
dependence, governed by the available resonant transitions and their
couplings to the pump pulse. The dependence on the pump-pulse duration
again follows the pattern seen in Fig.~\ref{fig:A2-tau}: shorter
pulses favor lower-$l$ processes, while longer pulses activate higher-$l$
channels, moderated by Rabi decay and inter-process competition.

Finally, Fig.~\ref{fig:A2-Enph} shows how the excitations depend
on both pump-pulse frequency and amplitude at fixed pulse duration.
As in previous plots, increasing $A_{0}$ enhances excitations, while
the frequency dependence reveals resonant structures that map out
the relevant band gaps and coupling strength to the pump pulse. The
photon-order resolved panels highlight how different absorption channels
contribute across the parameter space. This detailed mapping provides
practical guidance for selecting pump-pulse parameters to target specific
excitation pathways in germanium or similar semiconductors.

\section{Conclusions}\label{sec:Conclusions}

In this work, we have presented a detailed theoretical and numerical
investigation of residual carrier excitations in germanium driven
by ultrafast optical pulses, using the Dynamical Projective Operatorial
Approach (DPOA). Our study systematically addressed the fundamental
questions of how many carriers are excited, at which energies they
reside, and how much energy is absorbed by the electronic system as
a function of pump-pulse parameters.

The energy distribution of residual excited-carrier density, $\rho^{\mathrm{res}}(\varepsilon)$,
clearly exhibits a separation between electron and hole branches by
multiples of the pump-pulse photon energy, confirming the dominance
of $l$-photon resonant processes. The resonance strength, quantified
by $w_{l}(\varepsilon_{\mathrm{gap}})$, effectively captures the
spectral selectivity of the Gaussian pump pulse.

The excitations show a strong dependence on pump-pulse duration ($\tau_{\mathrm{pu}}$)
and amplitude ($A_{0}$). For high intensities, a non-monotonic, Rabi-like
behavior emerges as a function of pulse duration, where excitations
can be driven back toward the ground state for sufficiently long pulses.
This highlights the importance of coherent light-matter interaction
beyond simple perturbation theory.

Multi-photon decomposition reveals a distinct hierarchy and competition
between absorption channels. In germanium, two-photon processes are
generally the most efficient pathway across a wide parameter range.
One-photon processes become significant at higher pump-pulse frequencies,
while three-photon contributions require both high amplitude and long
pulse duration to activate, underscoring the role of effective Rabi
frequencies and state depletion.

The interplay between pump-pulse frequency and the electronic band
structure leads to a rich, non-monotonic dependence of total residual
excited-carrier density and absorbed energy on $\hbar\omega_{\mathrm{pu}}$.
Peaks and dips correspond to optimal (or suboptimal) matching with
specific band gaps and field coupling strength across the Brillouin
zone.

The parameter-space maps of $\Delta N^{\mathrm{res}}$ and $\mathcal{E}^{\mathrm{res}}$
provide a practical guide for tailoring ultrafast excitations. For
applications requiring high carrier density with moderate energy input,
the regime dominated by two-photon absorption is optimal. In contrast,
targeting higher-energy excitations via three-photon processes requires
carefully balancing pulse intensity and duration.

Our results explore the parameter-dependent landscape of ultrafast
optical excitations in realistic multi-band materials. The insights
gained here---particularly regarding the competition and activation
conditions for different multi-photon orders---are not specific to
germanium but represent general principles applicable to a wide class
of semiconductors. This work establishes a foundation for the rational
design of pump pulses to achieve desired electronic population distributions,
a crucial step toward controlling ultrafast phenomena for next-generation
optoelectronic and quantum devices.
\begin{acknowledgments}
The authors thank Matteo Lucchini for insightful discussions. The
authors acknowledge support by MUR under Project PNRR MUR Missione
4 (SPOKE 2) TOPQIN ``TOPological Qubit In driveN and reconfigurable
heterostructures''.
\end{acknowledgments}

 \bibliographystyle{apsrev4-2}
\bibliography{biblio}

\end{document}